\documentstyle[aps,prd,preprint,epsf]{revtex}
\begin{document}
\draft

\title{Bound states due to an accelerated mirror}

\author{Alberto Saa}
\address{Departamento de Matem\'atica Aplicada\\
IMECC -- UNICAMP, C.P. 6065,\\
 13081-970 Campinas, SP, Brazil}
\author{Marcelo Schiffer\thanks{On leave of absence of
Departamento de Matem\'atica Aplicada, UNICAMP, Campinas, Brazil}}
\address{Racah Institute of Physics \\
Hebrew University of Jerusalem\\
91904 Jerusalem, Israel}

\maketitle

\begin{abstract}
In this Brief Report we discuss an effect of accelerated mirrors
 which remained hitherto unnoticed, 
the formation of a field condensate near its surface for massive
fields.  From the view point
of an observer attached to the mirror, this is effect is rather natural
because a gravitational field  is felt there. The novelty here 
is that since the effect is not observer dependent even inertial
 observers will detect the formation of this condensate.
We further show that  this localization is in agreement with
Bekenstein's entropy bound.

\end{abstract}


The study of moving mirrors has been highly valuable to the
understanding of accelerating system physics and of quantum field
 theory on curved spacetimes\cite{BD,BMPS}. The case of a two
 dimensional massless scalar
field in the presence of a moving mirror has enlightened the
 process of particle creation due to gravitational fields, especially the
Hawking radiation. There is a vast literature on the subject 
and we indicate \cite{BD,BMPS} for references.

The case of massless fields is particularly convenient because of the
facility to impose the boundary condition on the mirror. The general
solution of the massless Klein-Gordon equation  $\Box\phi=0$ has the
form $\phi=f(v)+g(u)$, with $v=t+x$, $u=t-x$, and arbitrary $f$ and $g$.
The reflection condition is that $\phi$ vanishes on the mirror.
Assuming the mirror trajectory to be expressed by $v=z(u)$, the
solution takes the simple form $\phi=f(v) - f(z(u))$, valid for any
trajectory. From this solution, one can read off the Doppler shift
associated with the mirror trajectory,  and by exploiting Bogoliubov 
transformations determine the particle content of the vacuum state 
in the presence of the mirror as seen from the usual Minkowski
vacuum.

In the present work we investigate the behavior of a massive scalar
field $\phi$ in the presence of a uniformly accelerated mirror in 
two dimensions. Our conclusions are rather surprising: we show that 
the mirror acceleration implies on the existence of bound states for
$\phi$, 
hence inducing spatial localization. We further  discuss  that
 quantum
states in the
presence of the accelerated mirror differ in an essential  way 
from the case of an inertial mirror. This can be  hinted  by 
because in the first situation the
energy spectrum is discrete while in the second  one it is 
continuous,  precluding the existence of  a  Bogoliubov 
transformation 
relating  the corresponding modes. From a more mathematical
standpoint, this is a consequence of the fact that the
former set of modes are squared integrable
while the latter are  not.

Let us consider a real massive scalar field  with mass $m$
confined to the right region of a uniformly accelerated mirror. 
The mirror's trajectory corresponds to the hyperbola $x^2 - t^2 =\ 
\rm const$, $x>0$. This situation is described by the solutions of 
the massive Klein-Gordon equation
\begin{equation}
\label{kg}
\left(\Box + m^2 \right) \phi =0,
\end{equation}
such that
$\phi=0$ on the mirror trajectory.  We introduce  Rindler
coordinates $t=y\sinh a \eta$ and $x = y\cosh a \eta$, covering the
region $x\ge |t|$ of the Minkowski spacetime.
In this coordinate system, the
trajectory of a uniformly accelerated mirror is given by the lines
of constant $y=y_0$. In particular, for $y=a^{-1}$ the proper
acceleration is equal to $a$. 
Equation (\ref{kg}) becomes
\begin{equation}
\left(\frac{1}{a^2} \frac{\partial^2}{\partial\eta^2} - y
\frac{\partial}{\partial y}
y \frac{\partial}{\partial y} + m^2 y^2
\right)\phi = 0.
\end{equation}
Writing $\phi = e^{i\omega\eta}f_\omega(y)$ one has the following
equation
for $f_\omega$
\begin{equation}
\label{b}
f''_\omega + \frac{f'_\omega}{y} + \left( \frac{\omega^2}{a^2 y^2} - m^2
\right) f_\omega = 0.
\end{equation}
Equation (\ref{b}) is the Bessel's equation with imaginary order
and argument. Its solutions have the form\cite{GR}
$f_\omega = A I_{i\frac{\omega}{a}}(my) + BK_{i\frac{\omega}{a}}
(my)$. Since $I_{i\frac{\omega}{a}}(my)$
goes as $e^{my}$ for $y\rightarrow \infty$ we set $A=0$ and $B$
 can
be determined by normalization. Imposing the boundary
condition $K_{i\frac{\omega}{a}}(my_0) = 0$ one gets 
an eigenvalue equation for $\omega$. We assume hereafter that
$y_0=a^{-1}$. The following
representation for the 
Bessel functions\cite{GR}
\begin{equation}
\label{repr}
K_{i\frac{\omega}{a}}(m a^{-1}) = 
\int_0^\infty e^{-\frac{m}{a} \cosh t} \cos \frac{\omega}{a} t \, dt.
\end{equation}
is convenient for our purposes. We can determine $\omega$ numerically
from (\ref{repr}). We could do
it with good accuracy to the first eigenvalues. 
Figure 1 shows the aspect of the first
eigenvectors $f_{\omega_n}$. For $\frac{\omega}{a} \gg 1$, one
 can integrate this expression in the saddle  point
approximation and obtain  the asymptotic localization of
the zeroes of $K_{i\frac{\omega}{a}}(ma^{-1})$.
Calling
\begin{equation}
f(t)=-\frac{m}{a}  \cosh t + i \frac{\omega}{a} t
\end{equation}
we obtain  from the stationary phase condition
\begin{equation}
e^{t_0} = i\frac{\omega}{m} \pm i 
\sqrt{\left( \frac{\omega}{m}\right)^2 -1} \, .
\end{equation}
In  the $\omega/m \gg 1$ limit there is one  root   at  $t_0
\rightarrow -\infty$, which is far away from the integration
domain and a second one located approximately at  $e^{t_0}
\approx 2i\omega/m$. Superimposing a branch cut at the negative  real
axis, we obtain at this approximation,
\begin{equation}
f(t_0)= i \frac{\omega}{a} \left( \log \frac{2 \omega}{m e} + i
\frac{\pi}{2} \right)
\end{equation}
Therefore,
\begin{equation}
K_{i\frac{\omega}{a}}(m a^{-1}) \approx  {\rm Re} \left( 
e^{-\frac{\pi \omega}{2 a}} e^{ i \frac{\omega}{a}\log \left( 
\frac{2 \omega}{m e}\right) + \phi} 
 \int  \exp \left(-i \frac{\omega}{2a} s^2 e^{2i \phi}\right) ds 
\right),
\end{equation}
where  $s e^{i \phi} = t-t_0$. The  appropriate  deformation of
 the contour integration yields for $\phi=-\pi/4$.
 Putting all these pieces together,
\begin{equation}
K_{i\frac{\omega}{a}}(m a^{-1}) \approx \sqrt{\frac{2 a
\pi}{\omega}} e^{-\omega \pi/2a} \cos\left[ \frac{\omega}{a}
\log \left(\frac{2 \omega}{m e}\right) - \frac{\pi}{4}\right].
\end{equation}
Defining $z=2 \omega/m e$, we obtain the asymptotic expression
for the field eigenvalues,
\begin{equation}
\label{disp}
z\log z = \frac{a}{em}\left(2n + \frac{1}{2}\right)\pi,
\end{equation}
where $z= \frac{2\omega}{e m}\gg 1$.

Returning to the usual coordinates, we have finally that the
solutions of the massive Klein-Gordon equation in the presence
of a uniformly accelerated mirror with proper acceleration
$a$ has the form
\begin{equation}
\label{acc}
\phi(t,x) =\ \sum_{n=0}^\infty a_n \, e^{-i\frac{\omega_n}{a} {\,\rm arctanh\,}
\frac{t}{x}}
K_{i\frac{\omega_n}{a}}\left(m\sqrt{x^2 - t^2} \right) + {\ \rm hermitean\
conjugated},
\end{equation}
where $\omega_n$ are the zeroes
 of $K_{i\frac{\omega}{a}}(m a^{-1})$. Note
that we have a discrete number of modes, 
in contrast to the case of a mirror at rest 
in $x=a^{-1}$, for which we have
\begin{equation}
\label{rest}
\phi(t,x) = \int  dk \, b_k \, e^{-i\sqrt{k^2 + m^2}t}\sin k(x-a^{-1}) +
 {\ \rm hermitean\ conjugated},
\end{equation}
and the modes are continuous. As we already mentioned, 
 the  usual Bogoliubov transformation connecting (\ref{acc}) 
and (\ref{rest}) cannot be constructed.

Let us now consider two physically inequivalent situations. First,
suppose that an uniformly accelerated mirror with proper
acceleration $a$
ceases its motion at $t=0$ and remains at rest for $x=a^{-1}$. For
$t<0$,
the scalar  field $\phi$ is described by the superposition of
modes given in  eq. (\ref{acc}) and is totally determined by $m$ 
and $a$.
As  long  as the mirror starts its inertial trajectory for $t>0$
the field is described by another superposition of modes,
namely, that one given by  eq.[\ref{rest}]. The
coefficients $b_k$ and $a_n$ are related by a singular (non-invertible)
Bogoliubov transformation. Such a transformation can be determined
by equaling (\ref{acc}) and (\ref{rest}) and their first time
derivatives for $t=0$ and using that
both sets of modes are orthonormal with respect to the usual time
independent inner product of the Klein-Gordon equation. We have
$b_k = \alpha_{kn} a_n + \beta_{kn}a^\dagger_n$, where
\begin{eqnarray}
\label{four}
\alpha_{kn} &=& \frac{1}{2\pi}\int_0^\infty dy 
K_{i\frac{\omega_n}{a}}(my + ma^{-1})\sin ky \left( 
1 + \frac{\omega_n}{\sqrt{k^2+m^2} \frac{1}{ay + 1}} 
\right),\nonumber \\
\beta_{kn}  &=& \frac{1}{2\pi}\int_0^\infty dy 
K_{i\frac{\omega_n}{a}}(my + ma^{-1})\sin ky \left( 
1 - \frac{\omega_n}{\sqrt{k^2+m^2} \frac{1}{ay + 1}} 
\right).
\end{eqnarray}
We have that an originally localized state at $t=0$ starts a 
spreading phase and for $t>0$ it becomes totally unlocalized.

The other relevant situation is that one for which the mirror originally
at rest in $x=a^{-1}$ starts a uniformly accelerated motion with
proper acceleration 
$a$ at $t=0$. In this case, the scalar field is in a
superposition of plane wave states for $t<0$, and we know that
 for
$t>0$ it will be certainly localized. However, we cannot determine
the coefficients $a_n$ from $b_k$ easily as in the previous case:
the situation depends on the derivatives of the mirror acceleration
at $t=0$. Nevertheless, we know that for large $t$, when the
 transient effects are no  longer appreciable the 
field will relax to a superposition of modes given
eq. (\ref{acc}). As the the mirror starts its motion the 
originally unlocalized scalar field will  be dragged to near the 
mirror. Clearly, transients will  be
present which depend  very much on the way the mirror is set
into accelerated motion.

This consideration suggests a very attractive scenario:  
 the eventuality of having Bose-condensation for a system
which is not confined inside a box, in which confinement is
produced quantum mechanically. Thus, let us consider a situation where
 prior to the acceleration of the mirror, the field is in
 a thermal state at some fixed temperature. As the mirror 
accelerates, the state becomes localized in the 
vicinity of the mirror's face. By the time transients get over 
 and relaxation takes place the quantum  state relaxes  to a 
maximum entropy
 configuration,  a thermal state  localized in a region  whose
typical length scale is $m^{-1}$ (this scale  arises from the
exponential factor  in the asymptotic expansion of the
 $K_{i\frac{\omega}{a}}$).  As well know in condensed matter physics,
the excited states has a finite storage capacity. The total number of quanta
that can be accommodated in the system is
\begin{equation}
N= \sum_{\omega} \frac{1}{\tau^{-1} e^{\beta \omega}-1}
\end{equation}
where $\beta = \frac{1}{kT}$ and 
$\tau=\exp\left(\frac{\mu}{kT}\right)$, 
where $\mu=\mu(T)$ stands for the chemical potential and
$T$ for the system's temperature after relaxation.
In order to sum this series we have to recall our dispersion (\ref{disp}) 
relation for the large eigenvalues. One can see that there is a no large
integer $l$ such that 
\begin{equation}
N \approx \frac{em}{2a}
\int_{l}^\infty \frac{d(z \log z)}{\tau^{-1} e^{\gamma z}-1} + 
\sum_{n=0}^l\frac{1}{\tau^{-1} e^{\beta \omega_n} -1}
\end{equation}
where $\gamma = \frac{em}{2kT}$. Calling $N_l$ the second term in the
previous equation, we obtain the maximum storage capacity in
the high excited states 
\begin{equation}
N_e \equiv N - N_l= \frac{em}{2a} \int_l^\infty
\frac{\tau d(z\log z)}{e^{\gamma z} -\tau}.
\end{equation}
Recalling that $0\le\tau\le 1$,   
the r.h.s. of this equation is clearly a monotonic 
function of $\tau$, and thus
\begin{equation}
N_e \leq \frac{em}{2a}\int_l^\infty \frac{d(z\log z)}{e^{\gamma z}-1}.
\end{equation}
This shows the finite storage capacity in the excited states. 
For $N \geq N_e$, boson condensation takes places and
a thin film  is formed on the face of the accelerated mirror.

In our model, the scalar field is selfconfined into a region of
typical length $R = m^{-1}$ near to the mirror surface. We could
ask about a possible violation of the
Bekenstein's bound on the entropy of localized systems \cite{bek}
\begin{equation}
\label{bou}
S \le 2\pi \frac{E R}{\hbar},
\end{equation}
where $S, E$ and $R$ are the system's entropy, energy and largest linear
dimension,
owing to the large density of states of our system  in the high
frequency region. In our case, 
$E$ and $R$ are measured in the accelerated frame.
A heuristic argumentation suggests that one does not have a violation
of (\ref{bou}). Consider a massive scalar field confined into a 
box of size $R$ under a given acceleration. 
The spectrum of high frequency modes (frequencies
much bigger than both $R^{-1}$ and $m^{-1}$) should be very close
to the spectrum of a masless field in similar conditions. But for the
massless case, one should not expect any violation of Bekenstein's
bound since one has thermal photons in an external field. We will
show bellow that this is indeed the case.

 An upper bound on  $N^*$, the number of possible configurations with a
given energy,
 is provided by \cite{review1,review2}
\begin{equation}
N^*(E) = \sum_{i=0}^\infty g_i N^*(E-\omega_i),\, 
\end{equation}
where $g_i$ is $i$-th's level degeneracy ($\omega_i$ is the corresponding
frequency).
With the  ansatz $N^*(E) = e^{\frac{\Gamma E}{\hbar}},\,\Gamma >0$,
\begin{equation}
S \leq \frac{\Gamma E}{\hbar} , 
\end{equation}
with $\Gamma$ given by the solution of the equation
\begin{equation}
\sum_{j=0}^\infty g_j e^{-\Gamma \frac{\omega_j}{\hbar}} =1 \, .
 \end{equation}
Because
\begin{equation}
z^2 \ge z\log z = \frac{a}{em}\left( 2n - \frac{1}{2}\right)\pi
\end{equation}
and $g_i = 1$ for our case, an estimative for 
an upper bound on $\Gamma$ follows from
\begin{equation}
\sum_{j=0}^\infty 
\exp\left(-\frac{\Gamma}{\hbar}
\sqrt{\frac{ema}{2}\left(2j-\frac{1}{2}\right) \pi} \right) = 1,
\end{equation}
which in the continuous limit provides 
\begin{equation}
\Gamma \approx (e m a\pi)^{-1/2}\hbar.
\end{equation}
Consequently,
\begin{equation}
S \leq  m^{-1} (\sqrt{\frac{m}{e \pi a}})  \frac{E}{\hbar}
\end{equation}

Recalling that the typical linear dimension of the system is $R 
\approx m^{-1}$, 
violations of the bound would occur  at scales $m \gg a$.
Nevertheless, at this mass scale the asymptotic expansion 
obtained for the eigenvalues is no longer valid  and the above
expression in not valid anymore.

\acknowledgments

The authors are grateful to Prof. J.D. Bekenstein for discussions
and FAPESP for the financial support.

\begin{figure}[p]
\hfill\hbox{\epsfxsize=10cm\epsfbox{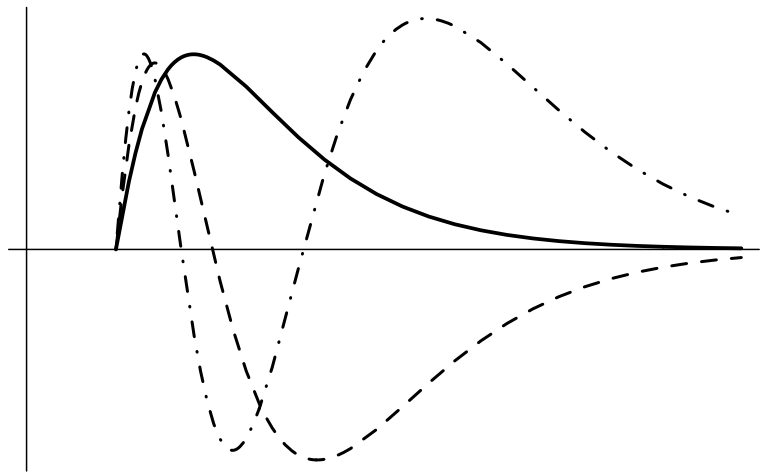}}\hfill\hfill
\caption{\label{fig} The first eigeinvectors $f_{\omega_n}$. The 
corresponding eigenvalues are 
$\omega_0 \approx 2.96 m$, $\omega_1 \approx 4.53 m$, 
and $\omega_2 \approx 5.88 m$ (assumed that $a=m$). }
\end{figure}

\end{document}